\documentclass[
    ,final            
  ]
  {aipproc}

\layoutstyle{6x9}
\usepackage{graphicx,color}

\usepackage{natbib}
\newcommand{\be}{\begin{eqnarray}}
\newcommand{\ee}{\end{eqnarray}}

\newcommand\simgreater{\,\lower0.7ex\hbox{$\stackrel{>}{\sim}$}\,}
\newcommand\simless{\,\lower0.7ex\hbox{$\stackrel{<}{\sim}$}\,}
\newcommand\msol{$M_\odot$}

\newcommand{\nue}{\ensuremath{\nu_{{\rm e}}}}
\newcommand{\nuebar}{\ensuremath{\bar{\nu}_{{\rm e}}}}

\newcommand{\nux}{\ensuremath{\nu_{{\rm x}}}}
\newcommand{\nuxbar}{\ensuremath{\bar{\nu}_{{\rm x}}}}
\newcommand{\num}{\ensuremath{\nu_{\mu}}}
\newcommand{\numbar}{\ensuremath{\bar{\nu}_{\mu}}}
\newcommand{\nut}{\ensuremath{\nu_{\tau}}}
\newcommand{\nutbar}{\ensuremath{\bar{\nu}_{\tau}}}

\begin{document}

\title{Mechanisms of Core-Collapse Supernovae \& Simulation Results from the CHIMERA Code}



\author{S. W. Bruenn}{
  address={Physics Department, Florida Atlantic University, 777 W. Glades Road, Boca Raton, FL 33431-0991}
}

\author{A. Mezzacappa}{
  address={Physics Division, Oak Ridge National Laboratory, Oak Ridge, TN 37831--6354}
}

\author{W. R. Hix}{
  address={Physics Division, Oak Ridge National Laboratory, Oak Ridge, TN 37831--6354}
}

\author{J. M. Blondin}{
  address={Department of Physics, North Carolina State University, Raleigh, NC 27695-8202}
}

\author{P.  Marronetti}{
  address={Physics Department, Florida Atlantic University, 777 W. Glades Road, Boca Raton, FL 33431-0991}
}

\author{O. E. B. Messer}{
  address={Center for Computational Sciences, Oak Ridge National Laboratory, Oak Ridge, TN 37831--6354}
}

\author{C. J. Dirk}{
  address={Physics Department, Florida Atlantic University, 777 W. Glades Road, Boca Raton, FL 33431-0991}
}

\author{S. Yoshida}{
  address={Max-Planck-Institut fur Gravitationsphysik, Albert Einstein Institut, 
Golm, Germany}
}

\begin{abstract}
Unraveling the mechanism for core-collapse supernova explosions is an outstanding computational challenge and the problem remains essentially unsolved despite more than four decades of effort. However, much progress in realistic modeling has occurred recently through the availability of multi-teraflop machines and the increasing sophistication of supernova codes. These improvements have led to some key insights which may clarify the picture in the not too distant future. Here we briefly review the current status of the three explosion mechanisms (acoustic, MHD, and neutrino heating) that are currently under active investigation, concentrating on the neutrino heating mechanism as the one most likely responsible for producing explosions from progenitors in the mass range $\sim$ 10 to $\sim$ 25 M$_{\odot}$. We then briefly describe the CHIMERA code, a supernova code we have developed to simulate core-collapse supernovae in 1, 2, and 3 spatial dimensions. We finally describe the results of an ongoing suite of 2D simulations initiated from a 12, 15, 20, and 25 M$_{\odot}$ progenitor. These have all exhibited explosions and are currently in the expanding phase with the shock at between 5,000 and 10,000 km. We finally very briefly describe an ongoing simulation in 3 spatial dimensions initiated from the 15 M$_{\odot}$ progenitor.
\end{abstract}

\maketitle

\section{Introduction}
\label{sec:Intro}

Much progress has been made unraveling the core collapse supernova mechanism in the past ten or so years, but it still remains an unsolved problem. It is apparent from observations that core-collapse supernovae exhibit large-scale anisotropies. Spectropolarimetry, the large average pulsar velocities, and the morphology of highly resolved images of SN 1987A all suggest that anisotropy not only develops, but likely develops very early on in the explosion \citep[e.g., see][for reviews and references]{arnett_bkw89, mccray_93, nomoto_skys94a}. This observed asymmetry is now believed to be a manifestation of the underlying essential multidimensional nature of the supernova mechanism. Realistic numerical core-collapse supernova modeling thus requires multidimensional techniques and correspondingly massive amounts of computer resources to capture this important macro-physics. Additionally, during and following core-collapse a plethora of microphysics comes into play and must be computed accurately as the supernova mechanism appears marginal. Inaccuracies in any part of a numerical simulation can prejudice the outcome. It is therefore not surprising that the supernova mechanism is taking a long time to unravel.

\section{Mechanisms}
\label{sec:Mech}

Three explosion mechanisms have been the focus of current research: (1) the acoustic mechanism, (2) the MHD mechanism, (3) and the neutrino heating mechanism. 

\subsection{Acoustic Mechanism}

The acoustic mechanism was discovered by \citet{burrows_ldom06} in the simulation of an 11 \msol\ progenitor. They found that long after shock stagnation ($\simgreater 0.6$ s post-bounce) turbulence and anisotropic accretion on the proto-neutron star excites and maintains vigorous g-mode oscillations which radiate intense sound waves, the energy coming from the gravitational binding energy of the accreted gas. As these sound waves propagate outward through the negative density gradient into the surroundings, they steepen into shocks and their energy and momentum are efficiently absorbed, powering up the supernova explosion. Thus, the proto-neutron star acts like a transducer converting the gravitational energy of infall into acoustic energy which propagates out and deposits energy in the surroundings, in analogy with the neutrino transport mechanism. They subsequently found \citep{burrows_ldom07}  that the acoustic mechanism is able to explode a variety of progenitors at late times ($\simgreater 0.6$ s). The physical reality of this mechanism is being debated as it has not been observed in simulations by other investigators when these simulations have been pushed to late times, although their numerical techniques, though different, are capable of capturing this phenomenon. A further note of caution is cast by a recent study by \citet{weinberg_q08} that finds that the damping of the primary mode by the parametric instability causes the primary l = 1 g-mode to saturate at an energy two orders of magnitude lower than that required to power a supernova. 

\subsection{MHD Mechanism}

Magnetic fields threading a progenitor are frozen in the gas on all relevant core-collapse time scales. On core collapse these magnetic fields will be amplified both by flux conservation during matter compression and by being wound up toroidally by the differential rotation of the core. Simulations with increasing sophistication have shown that if the iron core before collapse is threaded by very strong magnetic fields (B $\ge$ 10$^{12}$ gauss), then this in combination with rapid rotation can produce jet-like explosions magnetically on a prompt time scale \citep{leblanc_w70, mueller_h79, symbalisty_84, takiwaki_kns04, yamada_s04, sawai_ky05, burrows_dlom07, dessart_blo07, mikami_smh08, sawai_ky08}. Furthermore, it has been recognized that initially weak magnetic fields can be amplified to equipartition values exponentially by the magnetorotational instability \citep{balbus_h91, akiyamawml03, shibata_lss06}. Notwithstanding all this, it must be appreciated that the maximum magnetic energy that can be achieved in a differentially rotating core is the free energy, $t_{\rm free}$, of the differential rotation, i.e., the difference between the energy of the differentially rotating core and the same core uniformly rotating with the same angular momentum, and

\begin{equation}
T_{\rm free} \le T_{\rm rot} = 4 \times 10^{51} \left( \frac{ \kappa_{\rm I} }{0.3} \right) \left( \frac{ M }{ 1.4 M_\odot} \right) \left( \frac{ R }{10 \mbox{ km}} \right)^{2} \left( \frac{ P_{\rm rot} }{2 \mbox{ ms} } \right)^{-2} \mbox{ ergs}
\label{eq:p1}
\end{equation}

\noindent \citep{shibata_lss06}. Thus rather small initial rotation periods, $\le 2 \mbox{ ms}$, for newly formed neutron stars are required if enough magnetic energy is to potentially arise to power up the typical supernova. These small rotation periods are at variance with the calculated rotational periods of the magnetized cores of supernova progenitors \citep{heger_ws05}, and the extrapolated periods of newly formed neutron stars ($\ge 10$ ms). Both of these constraints are ``soft'' (stellar evolutionary calculations with rotation and magnetic fields are not ab initio, and we have not yet observed a newly formed neutron star), but if they hold then the MHD mechanism will only be relevant to a subset of core collapse supernovae. However, the observations of magnetars, long-duration gamma-ray bursts, and hints of highly collimated material in some supernova remnants suggests a subclass of events that are magnetically driven.

\subsection{Neutrino Heating Mechanism}

\begin{figure}[!h]
\vspace{0.cm}
\hspace{1.cm}
\setlength{\unitlength}{1.0cm}
{\includegraphics[width=11.0 cm]{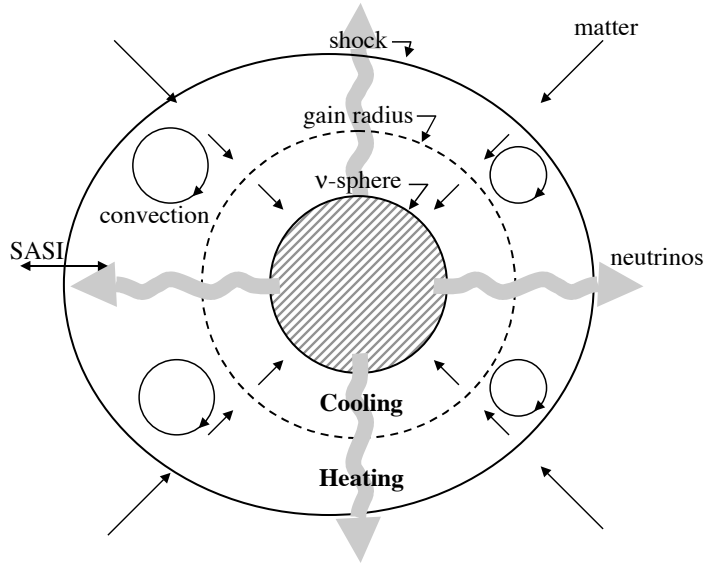}}
\caption{\label{Profile}
Entropy and velocity configuration snapshots of the Model 11M\_Sym\_32R.}
\end{figure}

The neutrino heating mechanism has a long pedigree extending back to the seminal paper of \citet{colgate_w66} and its more modern incarnation \citep{bethe_w85}. Following the collapse and bounce of the inner core of massive star at the endpoint of its normal thermonuclear evolution, the shock launched at core bounce stalls in the outer core, losing energy (and therefore post-shock pressure) to nuclear dissociation and electron neutrino losses. Within a short time ($\sim$ 50 ms) a thermodynamic profile is established as illustrated in Figure \ref{Profile} in which infalling matter encountering the outward flow of neutrinos undergoes net heating between the shock and the so-called gain radius, and net cooling below, due to the different neutrino heating and cooling radial profiles. Crudely speaking, for neutrino heating to be successful in powering an explosion a fluid element must be heated sufficiently while it resides in the heating layer to reenergize the shock. Sophisticated analyses \citep{burrows_g93, janka_01} have shown that the criteria for generating explosions in 1-D by neutrino heating depends on both the efficiency of heating between the gain radius and the shock and of cooling between the neutrinosphere and the gain radius.
 
Energy deposition by neutrinos plays the primary role in the neutrino heating mechanism, and the rate of energy deposition per nucleon, $\dot{q}$, can be written as

\begin{equation}
\dot{q} = \frac{X_{\rm n}}{ \lambda_{\nu_{\rm e}}^{a}} \frac{ L_{\nu_{\rm e}} }{4 \pi r^{2} } \langle \epsilon_{\nu_{\rm e}}^{2} \rangle \frac{1}{f_{\nu_{\rm e}}} + \frac{X_{\rm p}}{ \lambda_{\bar{\nu}_{\rm e}}^{a}} \frac{ L_{\bar{\nu}_{\rm e}} }{4 \pi r^{2} } \langle \epsilon_{\bar{\nu}_{\rm e}}^{2} \rangle \frac{1}{f_{\bar{\nu}_{\rm e}}},
\label{eq:p2}
\end{equation}

\noindent where the first and second terms express the absorption of electron neutrinos ($\nu_{\rm e}$'s) and antineutrinos ($\bar{\nu}_{\rm e}$'s), respectively. For the $\nu_{\rm e}$'s ($\bar{\nu}_{\rm e}$'s), $L_{\nu_{\rm e}}$ ($L_{\bar{\nu}_{\rm e}}$) is their luminosity, $\langle \epsilon_{\nu_{\rm e}}^{2} \rangle$ ($\langle \epsilon_{\bar{\nu}_{\rm e}}^{2} \rangle$) their mean square energy, and  and $\frac{1}{f_{\nu_{\rm e}}}$ ($\frac{1}{f_{\bar{\nu}_{\rm e}}}$) their inverse flux factor, which is a measure of their anisotropy. Clearly, an accurate calculation of  $\dot{q}$ requires an accurate calculation of both the energy spectrum and the angular distribution of the neutrinos.

\section{Status of the Neutrino Heating mechanism}

Simulations of core-collapse supernovae in spherical symmetry with considerable realism have been performed with Boltzmann neutrino transport, state-of-the-art neutrino interactions, and  with/without general relativity \citep{mezzacappa_lmhtb01, thompson_bp03, liebendorfer_mtmhb01, liebendorfer_rjm05}. These have not yielded explosions. Something is clearly missing.

An insight as to what the missing ingredient might be was developed during the 1990's and is the essential role played by multidimensional effects. Analyses of immediate post-bounce core profiles given by computer simulations had for a long time indicated that a variety of fluid instabilities are present \citep{epstein_79, arnett_87a, burrows_87, bethe_90}. The most important of these for the neutrino heating mechanism is the neutrino heating above the neutrinosphere. Because neutrinos heat the bottom o f the heating layer most intensely a negative entropy gradient builds up which renders the layer convectively unstable. In order for convection to grow, however, the fluid must remain in the heating layer for a critical length of  time; roughly the ratio of the advective timescale to some averaged timescale of convective growth timescale must be $\stackrel{\textstyle>}{\sim}$ 3 \citep{foglizzo_sj06}. If convection can get established in the hearing layer, hot gas from the neutrino-heating region will be transported directly to the shock, while downflows simultaneously will carry cold, accreted matter to the layer of strongest neutrino heating where a part of this gas, being cold, readily absorbs more energy from the neutrinos. The loss of energy accompanying the advection of matter through the gain radius is thereby reduced and more energy stays in the heating layer. It has been shown that convection in the heating layer will lead to explosions where the same model with the same neutrino luminosities and {\small RMS} energies fail when computed in spherical symmetry \citep{janka_m96}.

Another important ingredient missing in spherical symmetry was pointed out by \citet{blondin_md03} who discovered that the stalled shock is subject to low-mode aspherical oscillations, which they referred to as the standing accretion shock instability or `SASI.' The cause of this instability is still being debated \citet{foglizzo_gsj07, scheck_jfk08} arguing for an advective-acoustic cycle and \citep{blondin_m06, blondin_s07} for a purely acoustic cycle. The development of the SASI leads to an enlargement of the heating layer in one region and its diminution in another (Figure \ref{Profile}). Where the heating region is enlarged, the efficiency of neutrino heating is enhanced and the onset of convection is made more favorable if it has not already commenced. Where the hearing region is constricted, conditions are favorable for the establishment of down-flows or return-flows for large-scale convection. It has further been pointed out and supported by 2D simulations with parameterized neutrino sources that the development of the SASI leads to the large asymmetries observed for SN 1987A and other supernovae, and might account for the large observed velocities of neutron stars \citep{kifonidis_psjm06, scheck_kjm06}.

Several groups, the Arizona-Jerusalem collaboration, the Garching group, and the FAU-NCSU-Oak Ridge collaboration have modeled core collapse in 2 spatial dimensions with spectral neutrino transport and the results so far do not exhibit convergence. The Arizona-Jeruseleum group have not seen neutrino driven explosions in any of their simulations \citep{burrows_ldom06, burrows_dlom07}. They apply 2D neutrino diffusion but neglect any energy-bin coupling and velocity corrections, and employ Newtonian gravity. The Garching group \citep{rampp_j02, buras_rjk06a, buras_rjk06b, marek_j07} apply a ``ray-by-ray-plus'' treatment of neutrino transport, where transport is calculated along radial rays, neglecting neutrino shear and nonradial fluxes, but including all other lateral effects. Transport is computed by solving the neutrino number, momentum, and energy equations closed by a variable Eddington factor computed from a simplified Boltzmann equation. The neutrino microphysics is state of the art. Gravity is computed using an approximate general relativistic potential for spherical gravity and Newtonian higher moments. The Garching group has evolved an 11.2  M$_{\odot}$ and a 15 M$_{\odot}$ progenitor and finds explosions commencing after rather long post bounce times, 220 ms and 620 ms respectively. Their simulations were not carried long enough to ascertain the explosion energies directly. Using the CHIMERA code, our group (the FAU-NCSU-Oak Ridge collaboration) has carried out simulations for a suite of four progenitors of MS masses 12 M$_{\odot}$, 15 M$_{\odot}$, 20 M$_{\odot}$,and 25 M$_{\odot}$ evolved to core collapse by \citet{woosley_h07}. We obtain explosions for all these models and find that they power-up earlier compared with the Garching simulations. Our simulations are ongoing with the models having currently been evolved to post bounce times of from 400 to 700 ms. The CHIMERA code and our results will be described below.

\section{The CHIMERA Code}

The CHIMERA code is designed to simulate core-collapse supernovae in 1, 2, and 3 spatial dimensions from the onset of collapse to the order of 1 sec post bounce given present day state-of-the-art computational resources, such as the Cray XT4. It conserves total energy (gravitational, internal, kinetic, and neutrino) to within $\pm$ 0.5 B. The code currently has three main components:  a hydro component, a neutrino transport component, and a nuclear reaction network component. In addition there is a Poisson solver for the gravitational potential and a sophisticated equation of state. A preliminary version of the code was briefly described in \citet{bruenn_dmhbhm06, mezzacappa_bbhb07}.

The hydrodynamics is evolved via a Lagrangian remap implementation of the  Piecewise Parabolic Method (PPM) \citep{colella_w84}. A moving radial grid option wherein the radial grid follows the average radial motion of the fluid makes it possible for the core infall phase to be followed with good resolution. Following bounce, an adaptive mesh redistribution algorithm keeps the radial grid between the core center and the shock structured so as to maintain approximately constant $\Delta \rho/\rho$. For 256 radial zones, this ensures that there are at least 15 radial zones per decade in density. For 512 radial zones (our higher resolution runs) at least 30 radial zones per decade in density are maintained. The equation of state (EOS) blend of \citet{lattimer_s91}, highly modified \citet{cooperstein_85a}, and other details are described in \citet{bruenn_dmhbhm06}. The algorithm described in \citep{sutherland_bb03}, modified for greater rubustness, was employed to stabilize shocks oriented along grid lines. The gravitational potential is solved by means of the Newtonian gravity spectral Poisson solver described in \citep{mueller_s95}, and modified for application to 3D. Gravity was computed as described above for the Garching group.

Neutrino transport is implemented by a ``ray-by-ray-plus'' approximation \citep[cf.][]{buras_rjk06a} whereby the lateral effects of neutrinos such as lateral pressure gradients (in optically thick conditions), neutrino advection, and velocity corrections are taken into account, but transport is performed only in the radial direction. Transport is computed by means of multigroup flux-limited diffusion with a sophisticated flux limiter that has been tuned to reproduce results of a general relativistic Boltzmann transport results to within a few percent \citep{liebendorfer_mbmbct04}. All O($v/c$) observer corrections have been included. The transport solver is fully implicit and solves for four neutrino flavors simultaneously (i.e., \nue's, \nuebar's, \num's and \nut's (collectively \nux's), and \numbar's and \nutbar's (collectively \nuxbar's)), allowing for neutrino-neutrino scattering and pair-exchange, and different $\nu$ and $\bar{\nu}$ opacities. State-of-the-art neutrino interactions are included with full energy dependences.

The nuclear composition in the non-NSE regions of these models is evolved by the thermonuclear reaction network of \citep{hix_t99a}.  This is a fully implicit general purpose reaction network, however in these models only reactions linking the 14 alpha nuclei from $^{4}$He to $^{60}$Zn are used. Data for these reactions is drawn from the REACLIB compilations \citep{rauscher_t00}. The nucleons have only very small abundances at any time and are included to make the NSE-nonNSE transition smoother. The iron-like nucleus is included to conserve charge in a freezeout occurring with an electron fraction below 0.5.

\section{2D SImulation Results}

\begin{figure}[!h]
\vspace*{0cm}
\setlength{\unitlength}{1.0cm}
\begin{minipage}[t]{2.9in}
\hspace*{-1.cm}
{\includegraphics[scale = 0.40]{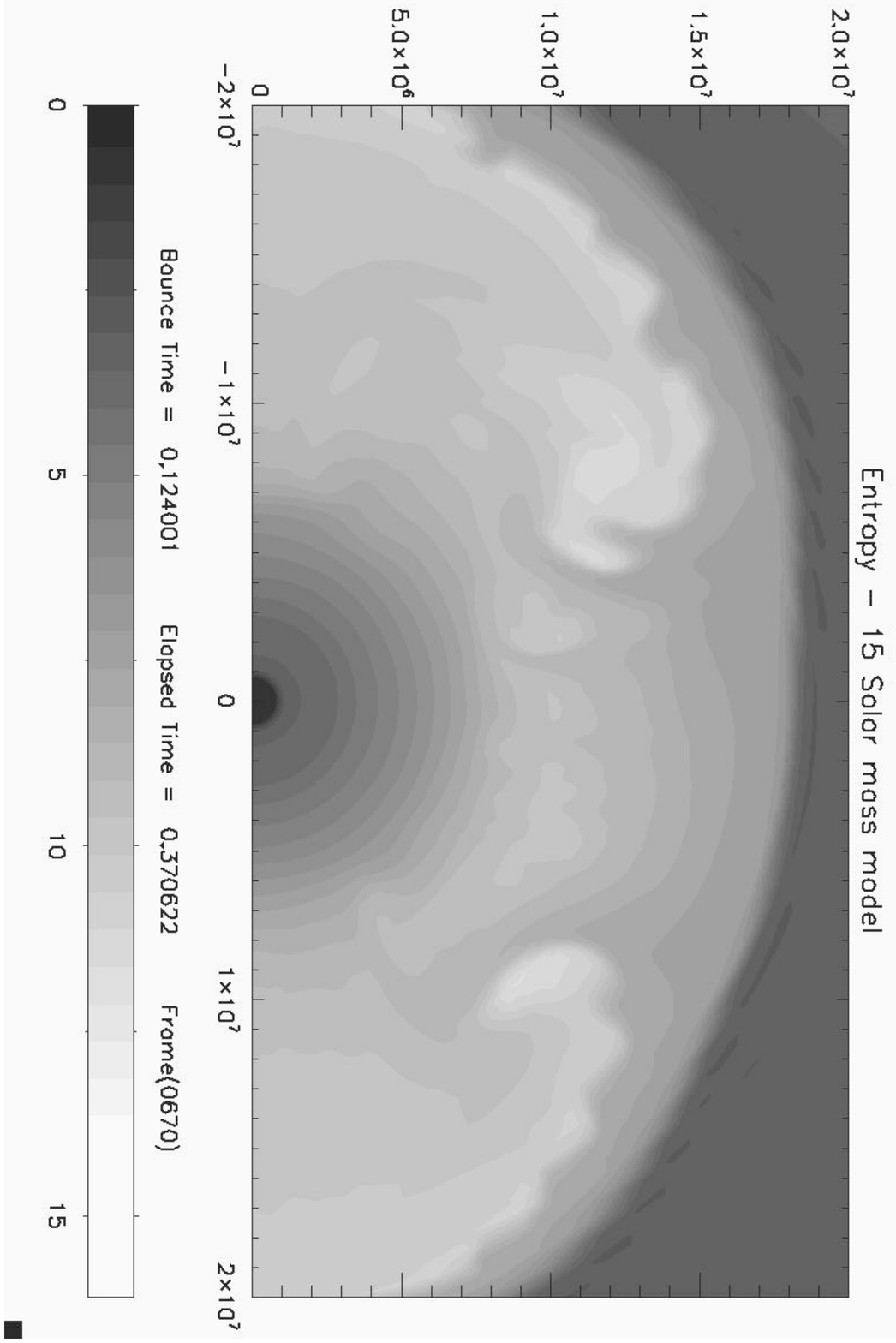}}
\vspace*{-1cm}
\caption{\label{fig:bwfig_1}
\small{}}
\end{minipage}
\hfill
\begin{minipage}[t]{2.9in}
\hspace*{-0.7cm}
{\includegraphics[scale = 0.40]{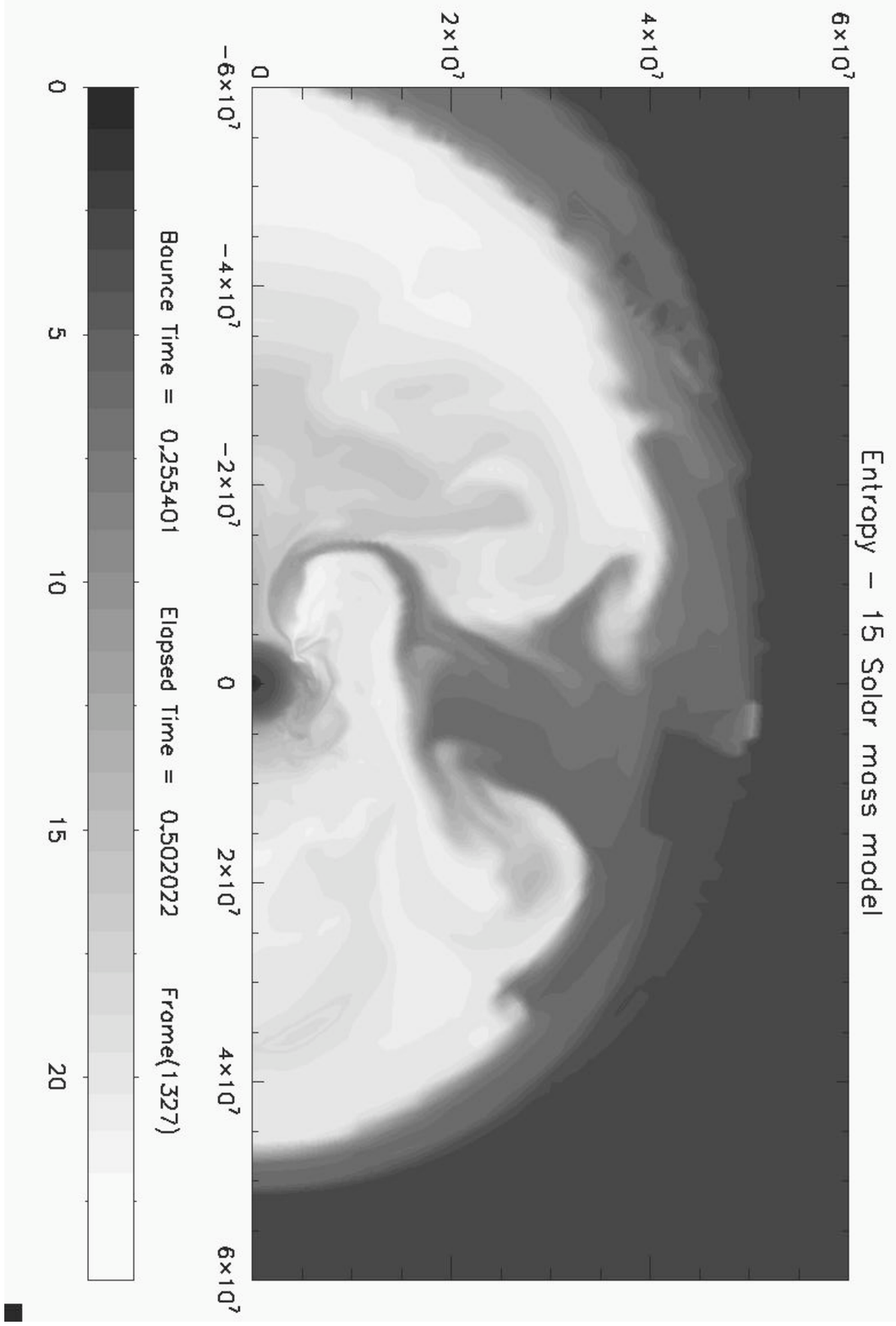}}
\vspace*{-1cm}
\caption{\label{fig:bwfig_2}
\small{Two snapshots of the entropy distribution in the 15 M$_{\odot}$ model at times 124 ms (left) and 255 ms (right) post-bounce. Note the difference in scale of the two figures,}}
\end{minipage}
\end{figure}

With the many improvements in the CHIMERA code and additions of more realistic physics (e.g., GR corrections, state-of-the-art neutrino microphysics) since our first results were reported in \citet{bruenn_dmhbhm06} we find that all our models explode as before, but now the nuclear energy released when the oxygen layers reach the shock does not play the critical role it did before. For a period of time after bounce (65 ms for the 12 and 15 M$_{\odot}$ models, 105 ms and 130 ms respectively for the 20 and 25 M$_{\odot}$ models) the 2D and corresponding spherically symmetric simulations track each other very closely. (This is apart from a brief episode in the 2D models of lepton driven convection in the proto-neutron star, followed by an equally brief episode of neutrino driven convection from the negative entropy gradient imprinted by the shock as it weakened.) During this period the heating layer becomes convectively unstable but its growth is suppressed in the 2D simulations because of the too rapid inflow of material. At the end of this period convective mushrooms begin to appear in the heating layer and at almost the same time the shock begins to exhibit SASI dipole and quadrupole deformations along the polar axis. This expands the heating layer in the polar regions and the enhanced convection in these regions quickly develops into a large-scale overturn (Fig. \ref{fig:bwfig_2}, left), with material rising and expanding in the polar regions and flowing towards the equatorial region. By 225 ms post-bounce for the 15 M$_{\odot}$ model (Fig. \ref{fig:bwfig_2}, right), as an example, large lobes of high-entropy neutrino heated gas are pushing the shock outwards in the polar regions while an equatorial accretion funnel has developed which is channeling lower-entropy newly shocked material down to the gain layer. The narrow end of the accretion funnel tends to oscillate slowly from one hemisphere to the other, seeming thereby to alternately pump up one hemisphere and then the other with energy by the enhanced neutrino emission at its base. Simultaneous with this convective activity is a large dipolar SASI oscillation of the shock on a more-or-less fixed quadrupolar background. It is likely that the lateral sloshing of material accompanying this SASI is what causes the base of the accretion funnel to oscillate between hemispheres, i.e. the dog is indeed wagging the tail rather than the other way around.

\begin{figure}[!h]
\vspace{0.cm}
\hspace{1.cm}
\setlength{\unitlength}{1.0cm}
{\includegraphics[width=11.0 cm]{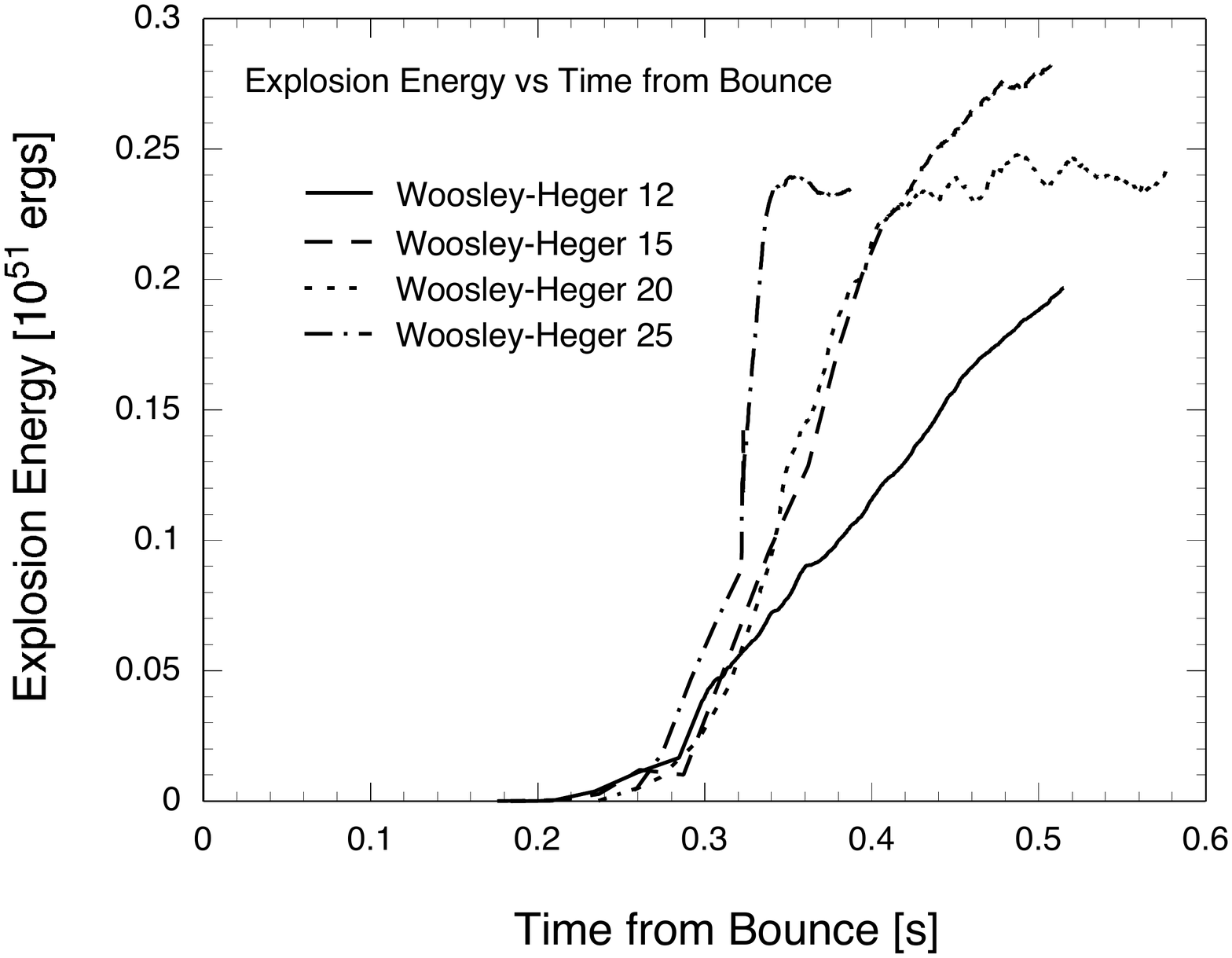}}
\caption{\label{Explosion}
Explosion energies as a function of post-bounce time.}
\end{figure}

Eventually a runaway condition is met and the shock begins to accelerate rapidly outwards. Measured by the first signs of positive radial post-shock velocities (a different criterion than that used by the Garching group) explosions commence at roughly 300 ms from bounce for the four models. The 12, 15, and 25 M$_{\odot}$ models exhibit a pronounced quadrupolar prolate configuration of the shock as the explosion commences, while the 20 M$_{\odot}$ the configuration of the shock is more dipolar. The ratio of the major to minor axis of the shock ranges from 1.5 (25 M$_{\odot}$ model) to 2.0 (15 M$_{\odot}$ model). The explosion energies, computed conservatively as the sum of the kinetic, internal thermal, and kinetic energies of outward moving material is shown in Figure \ref{Explosion}.

\section{3D SImulation Results}

We mention finally an ongoing medium resolution simulation in 3 spatial dimensions with 304 radial zones, 76 angular, and 152 azimuthal zones, initiated from the 15 M$_{\odot}$ progenitor. The simulation has progressed to 139 ms post-bounce and is exhibiting large scale convection patterns. We anticipate that this simulation will tell us much about the limitations of modeling core collapse in 2D.

\begin{theacknowledgments}
The authors would like to acknowledge partial funding by a grant from the DOE Office of Science Scientific Discovery through Advanced Computing (SciDAC) Program. S.W.B., P.M., and O.E.B.M. acknowledge partial support from an NSF-OCI-0749204 award. S.W.B., P.M., O.E.B.M., W.R.H., and A.M. acknowledge partial support from NASA award 07-ATFP07-0011. S.W.B., P.M., O.E.B.M., W.R.H., and A.M. also acknowledge a TACC Ranger award (TG-MCA08X010) and a DOE INCITE award at the National Center for Computational Sciences at the Oak Ridge National Laboratory (ORNL).  A.M., W.R.H., and O.E.B.M. are supported at ORNL, managed by UT-Battelle,  LLC, for the U.S. Department of Energy under contract DE-AC05-00OR22725.
\end{theacknowledgments}

\bibliography{BibTeX_list}

\end{document}